\def\be{\begin{equation}}
\def\ee{\end{equation}}
\def\ba{\begin{eqnarray}}
\def\ea{\end{eqnarray}}
 
\documentstyle[psfig]{elsart}
\begin{document}
\begin{frontmatter}
\psfigurepath{./}

\title{ On the Special Role of Symmetric Periodic Orbits in a 
Chaotic System}
\author[LCuernF]{L. Benet},
\author[LCuernM]{C. Jung},
\author[UWash]{T. Papenbrock},
\author[LCuernM]{T. H. Seligman}

\address[LCuernF]{Inst. de F\'{\i}sica,
National University of Mexico (U.N.A.M.), Cuernavaca, M\'exico}
\address[LCuernM]{Inst. de Matem\'aticas, 
National University of Mexico (U.N.A.M.), Cuernavaca, M\'exico}
\address[UWash]{Inst. for Nuclear Theory, University of Washington, 
Seattle, USA}

\begin{abstract}
After early work of H\'enon it has become folk knowledge that  
symmetric periodic orbits are of particular importance. 
We reinforce this belief by additional studies and we further 
find that invariant closed symplectic submanifolds caused by 
discrete symmetries prove to be an important complement to the 
long known role of the orbits. The latter have particular 
importance in semi-classics.
Based on the structural stability of hyperbolic horseshoes we give 
an argument that opens an avenue to the understanding of these facts.
\end{abstract}
\end{frontmatter}

\section{Introduction}

In 1965~\cite{Henon1} H\'enon charts the simple symmetric periodic 
orbits of the so-called Copenhagen problem, i.e. the planar reduced 
circular three-body problem with equal masses for the two heavy bodies
or, in other words, the planar orbits of a small planet or comet in a 
circular double star system. 
He states that the importance of these orbits resides in the fact that
in a subsequent paper~\cite{Henon3}, where he analyzes the general 
periodic orbits of the same system, he can discuss the latter 
essentially in terms of the former.

More recently we find a discussion of symmetric short periodic orbits 
in simpler systems~\cite{richter}, and in some sense it seems to be folk 
knowledge that such orbits are not only easy to find but usually also 
particularly relevant. 
In this context it is important to state what we mean by relevance.
This can not be a strict mathematical concept, but we may at least give 
one family of physical and one of mathematical examples, where this 
relevance is obvious. 

In semi-classical considerations of chaotic systems trace-formulae of 
the Selberg-Gutzwiller type~\cite{Gutz} play a central role. 
These formulae are sums over periodic orbits and from their structure 
it is clear that the leading terms are associated with short periodic 
orbits that are not too unstable. 
It is precisely such periodic orbits that according to~\cite{richter} 
may be readily found among the symmetric ones and specific calculations 
have often confirmed this as may be exemplified by the collinear states 
of Helium~\cite{Wint}.
Particular relevant work in this context was presented by Cvitanovic
and Eckhardt on the symmetry decomposition of the cycle expansion
for the zeta functions~\cite{cvit}.

On the other hand the discussion of the chaotic saddle of scattering 
problems in two dimensions implies a Smale horseshoe construction 
starting from a few fundamental periodic orbits which overshadow
the entire saddle~\cite{JungRuck1}. 
Again experience shows that in presence of symmetry, the symmetric 
orbits seem to be favored as fundamental periodic orbits. 

Recently we came across the importance of symmetric orbits in quite 
different contexts and the relevance is marked to an extent that can no 
longer be assumed to be casual. 
The purpose of this presentation is not only to show the essential role 
symmetric orbits play in certain systems. 
We shall show that their role subsists under significant deformations 
of the systems where the symmetry is destroyed. This should provide a 
clue as to the reason of their importance.

We shall start with an analysis of the chaotic saddle of the circular 
reduced three-body problem that shows that in this complicated case the 
infinite order horseshoe is well described by symmetric basic orbits 
though the parabolic ones (orbits that are asymptotically parabolae, 
not marginally stable orbits) have to be included~\cite{celmec1,celmec2},
thus extending the findings of H\'enon to the full scattering problem.
Next we shall consider results obtained for systems of many identical 
particles. 
We find short periodic orbits that are not only invariant under some 
subgroup of the symmetry group, but are mapped pointwise on themselves. 
Such periodic orbits and indeed the entire invariant submanifolds in 
which they lie are shown to be fundamental.
Finally we shall return to very simple systems such as the three-disc 
problem where we shall see that due to structural stability the relevance 
of such symmetric orbits persists under deformations, where the symmetry
is largely destroyed. 
This gives us a first hint as to the reason for the recurrent importance 
of these orbits. 

\section{Scattering of a comet from a double star}

Consider the coplanar scattering motion of a small body or comet off a 
binary star system on a circular orbit. 
Due to the huge ratio between the mass of the stars and that of the 
comet we can consider this to be a reduced three-body problem where we 
neglect the influence of the comet on the motion of the stars.
The Hamiltonian of this problem in rotating coordinates does not depend 
explicitly on time and therefore it is a constant of the motion, the 
Jacobi integral, which is given by:
\be
J={1 \over 2}(p_{x}^2 +p_{y}^2)-\omega(xp_{y}-yp_{x}) + V_{g}(x,y),
\label{eqone}
\ee
where the gravitational potential has the form
\be
V_{g}(x,y) = -{Gm_{1}\over \sqrt{(x-x_{1})^2+y^2}}-
{Gm_{2}\over \sqrt{(x-x_{2})^2+y^2}}.
\label{eqtwo}
\ee
Here, $m_1$ and $m_2$ are the masses of the stars, D their separation and 
$x_1=D m_2/M$ and $x_2=- D m_1/M$ their positions, with their center of
mass at the origin. 
$\omega$ is the angular velocity of their circular motion. 
One of the best explored systems is the so-called Copenhagen problem, 
where the two stars have equal masses.

\begin{figure}
\noindent\centerline{
\psfig{figure=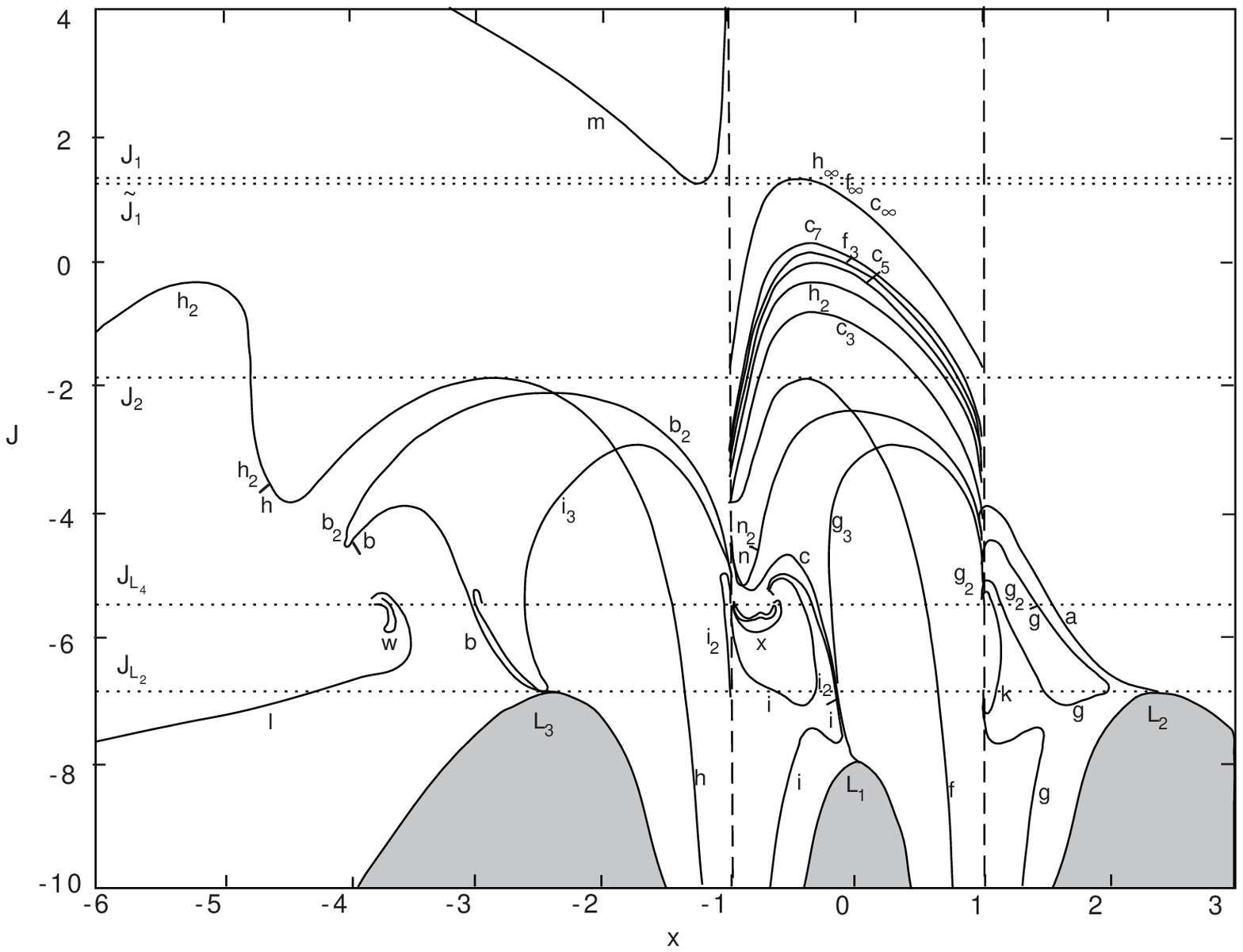,height=8cm}}
\caption{
H\'enons general chart of symmetric periodic orbits~\cite{Henon1}
(reproduced with permission from the author).
The vertical lines represent the position of the stars and the shadowed 
regions are the potential barriers defined by the Hill curves. 
The horizontal lines represent important values of the Jacobi 
integral where the main bifurcations take place.}
\label{figone}
\end{figure}

For the general case (different masses of the stars) it follows from 
Eqs.~(\ref{eqone}-\ref{eqtwo}) that the system has a symmetry resulting 
from simultaneous reflection with respect to the the x-axis and 
time reversal.
For the Copenhagen problem H\'enon~\cite{Henon1} explored the periodic 
orbits symmetric under this operation, focusing on the ones which crossed 
the x-axis only once with $ \dot y>0$. 
These he called the {\it simple symmetric periodic orbits}. 
He soon discovered that some related families of double and higher order 
symmetric periodic orbits also entered the picture in a relevant way,
and included them on the chart.
These new families of orbits have all many crossings with the x-axis 
($\dot y >0$), but only one or at most two perpendicular ones 
respectively for orbits with an odd or even number of crossings.
Figure~\ref{figone} shows H\'enons chart of simple symmetric periodic 
orbits for the Copenhagen problem.

This chart shows several regions with very distinct behaviour as 
we vary the Jacobi integral~\cite{celmec1}.
In particular, for the Jacobi integral interval defined by $J_1$ and
$J_2$ in Fig.~\ref{figone}, we find the families $h_n$, $c_n$ and 
$f_n$ of simple symmetric periodic orbits (the $c_n$ and $f_n$ 
families are symmetry related by the reflection on the y-axis 
and time reversal).
The subscript $n$ denotes the number of x-axis crossings with
$\dot y >0$ before the periodic orbit is closed.
Increasing values of $n$ imply longer periodic orbits and farther
position of the turning points.
In the Jacobi interval where we have the families $h_ \infty$ and 
$c_\infty$, it is not enough to include all the finite symmetric
periodic orbits, but we have to include also their accumulation 
points in our considerations. 
The latter are related to symmetric parabolic orbits that reach out 
to infinity; this implies that the chaotic saddle also reaches out to 
infinity which is no surprise because of the long-range nature of 
the Coulomb potential.

\begin{figure}
\noindent\centerline{
\psfig{figure=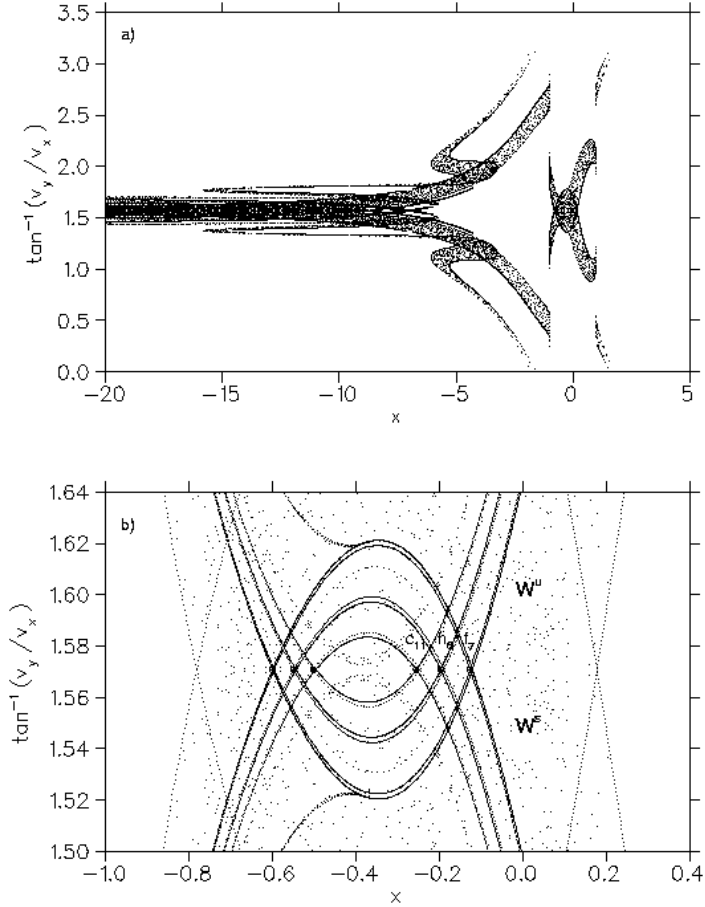,height=12cm}}
\caption{
Surface of section of the Copenhagen problem for $J=0.5$.
Only the manifolds of the hyperbolic and inverse hyperbolic fixed points 
associated with the periodic orbits $c_{11}$, $h_6$ and $f_7$ have 
been plotted.
(a)~Global structure of the stable and unstable manifolds. 
(b)~Enlargement of the region between the stars. 
The outer manifolds belong to parabolic orbits ($c_\infty$, 
$h_\infty$ and $f_\infty$).}
\label{figtwo}
\end{figure}

We shall demonstrate this statement for the value $J=0.5$ of the Jacobi 
integral which belongs to the above mentioned interval in H\'enons chart.
The other regions are discussed in detail in~\cite{celmec1}.
In Fig.~\ref{figtwo}a we present a Poincar\'e section constructed only
with the shortest symmetric periodic orbits; Fig.~\ref{figtwo}b is an 
enlargement of the region between the stars. 
The surface of section is defined by the position on the x-axis 
where the comet crosses this axis ($\dot y>0$), and by the angle 
$\gamma= tan ^{-1}(v_y/v_x)$ that the velocity vector forms in the 
synodic frame with the x-axis.

Notice first that the manifolds of the shortest periodic orbits
lie deep inside the corresponding horseshoe construction, and that
they appear within some external manifolds. 
It is known that the most important manifolds in the Smale horseshoe 
construction are precisely the outer ones, which build the so-called 
fundamental (curvilinear) rectangle~\cite{JungRuck1}.
By combining the arrangement of symmetric periodic orbits of 
Fig.~\ref{figone} with the structure of the surface of section of 
Fig.~\ref{figtwo}b one is led to conclude that the outermost manifolds 
of the present chaotic saddle correspond precisely to those of the 
accumulation points, i.e. to the limiting parabolic orbits.
Thus, to have a full understanding of the structure of the chaotic 
saddle for this case, beside the symmetric periodic orbits, one has to 
consider in the description the limiting parabolic orbits as part of 
the set of primitive periodic orbits.
Moreover, since the number of independent symmetric periodic orbits 
needed for the whole construction of this saddle is infinite, we are 
in the case of an infinite order horseshoe.

A similar argument can be made in the case of small mass ratio between 
the less massive star and the total mass of the binary system. 
One starts by analyzing the symmetric collision orbits one obtains 
assuming zero mass for the smaller star but considering its collisions 
with the comet. 
In this case no chaotic scattering occurs since no homoclinic or 
heteroclinic connections between such orbits exist. 
Yet it turns out that the symmetric periodic collision orbits dominate 
the situation for small but non-zero mass ratios where homoclinic and 
heteroclinic connections between such orbits do occur. 
Details of this more complicated situations are forthcoming 
in~\cite{celmec2}. 
In all the cases discussed above and in 
refs.~\cite{Henon1,celmec1,celmec2} the dominant role of the symmetric 
periodic orbits is manifest.

To conclude this section, we shall briefly mention another interesting 
case in celestial mechanics where the symmetry of the periodic orbits 
is crucial, the central configurations.
A central configurations is defined in the center of mass frame as one 
where the net force acting on all particles at a given time is radial 
and proportional to their distance from the origin~\cite{centralconf}.
Note that the constant of proportionality may be time dependent.
In this case, the interest is focused on constraining the initial
conditions in such a way that the particles stay for all times within
the submanifold spanned by the central configurations. 
These situations are important in the analysis of collision orbits 
and of expanding gravitational systems, among others~\cite{centralconf}.
Notice that this case is different from the situation discussed above, 
where the complete periodic orbits maps onto itself by the symmetry 
operation, whereas now each point of such an orbit maps onto itself.
Such configurations are a special case of the situation discussed in 
the next section.

\section{Closed invariant submanifolds in interacting few--body systems}

The central configurations discussed at the end of the previous 
section constitute a particular case of a larger family of closed 
invariant subspaces resulting from symmetries of the system. 
In our discussion we wish to exclude the well-known symmetry reduction 
of the dimensionality of a system in presence of a continuous symmetry 
that gives rise to additional constants of motion that are in involution. 
Rather we are interested in the following situation: 
The discrete symmetry group of a system or a discrete subgroup of a 
symmetry group may act trivially on a submanifold of phase space. 
Such a submanifold is invariant under the Hamiltonian flow of the system. 
The simplest example thereof is a reflection symmetry on a line or plane 
in configuration space.
In this case the two or four dimensional phase space generated by points 
on this line or plane and momenta along this line or in this plane 
constitute the submanifold. 
Any periodic orbit that exists in the submanifold is obviously symmetric 
and indeed trivially so as the symmetry maps the orbit pointwise onto 
itself.
The Hamiltonian may be restricted to such submanifolds and we can find 
periodic orbits in a comparatively low-dimensional space. 
This is a technical advantage, but their seems to be more to these 
invariant submanifolds. 
Prosen recently showed that such symmetry planes may carry scars of 
the wave-functions~\cite{Prosen97}.

In rotationally invariant systems of identical particles such 
invariant submanifolds occur systematically~\cite{PaSel}.
Indeed they occur whenever we have a direct product of two symmetry 
groups that have a common subgroup on a submanifold. 
In this case we can use the elements of one group to annihilate the 
action of those of the other group. 
The most trivial example is obtained by arranging the $n$ particles on a 
ring at equal distances and with momenta perpendicular to the ring. 
Not only are the $n$-cycles of the permutation group ${\mathrm S}_n $ 
and the rotations ${\mathrm C}_n$ isomorphic, but they act in the same
way on such a configuration. 
We can therefore consider a direct product 
${\mathrm C}_n \times {\mathrm C}_n $ where the first represents the 
permutational n-cycles and the second the corresponding inverse rotation 
about an axis perpendicular to the ring in its center. 
While this group acts non-trivially on the entire phase space it clearly 
acts trivially on the phase space generated by such ring-configurations. 
Note that the original phase space was 4-n dimensional while the 
submanifold is only 2-dimensional. 
More complicated configurations leading to smaller invariance groups and 
thus larger dimensional manifolds can be constructed~\cite{PaSel}.

An interesting question here will be to analyze the stability of such 
periodic orbits in the manifold and perpendicular to it. 
A number of examples have been studied~\cite{PaSel,PSW} and a wide 
variety of behaviour has been found in systems from three to nine 
particles in central harmonic and anharmonic fields interacting through 
Coulomb and quartic interactions. 
The small number of degrees of freedom does not imply that only a few 
particles are moving; rather it implies some form of collective movement 
of all particles. 
We shall show that this collectivity can have a physical meaning if the 
instability perpendicular to the submanifold is fairly small.

The obvious application derives from semi-classical considerations 
in quantum mechanics.
Semi-classical periodic orbit theory, as developed by Balian, Bloch,
Gutz\-willer and Berry~\cite{Gutz,potheory}, has been used successfully 
to analyze chaotic systems with few degrees of 
freedom~\cite{Wint,fewdegrees}.
Within the semi-classical approximation, the level density is given in 
terms of all periodic orbits of the classical system
\be
\rho(E)=\sum_{p}A_p(E)\exp{\frac{i}{\hbar}S_p(E)}.
\label{pos}
\ee
Here the amplitude $A$ depends mainly on the stability of the orbit 
labeled by~$p$, and $S_p$ is its action. 
Including only the shortest orbits in the sum~(\ref{pos}) yields a coarse 
grained spectrum, and therefore the long range correlations in the density 
of states are determined by the shortest and most stable orbits. 
Generically, individual periodic orbits are not associated with individual 
energy levels or wave functions. 
However, short and fairly stable periodic orbits may scar individual wave 
functions, i.e. a wave function displays higher than average intensities 
in the vicinity of a particular periodic orbit~\cite{scars}.

\begin{figure}[htb]
\noindent\centerline{
\psfig{figure=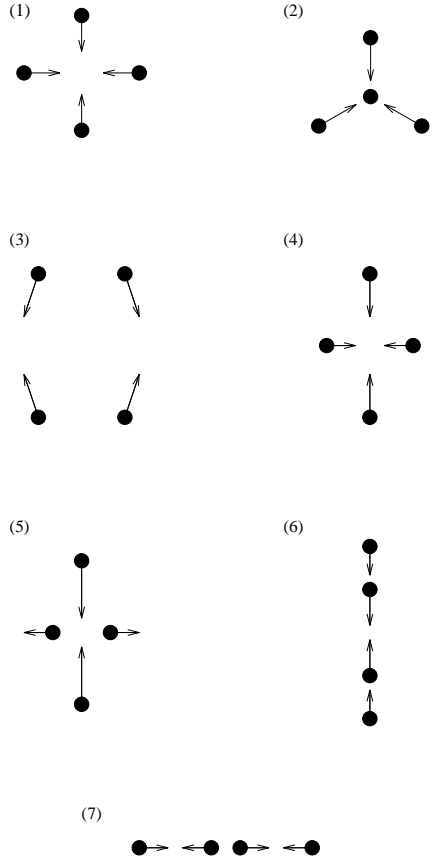,width=0.4\textwidth,angle=0}}
\caption{Schematic plots of configurations that restrict the classical 
motion to invariant submanifolds in the case of a 4--particle system. 
Positions and momenta of the particles are indicated by points and 
arrows, respectively.}
\label{figure1}
\end{figure}

In chaotic systems of two degrees of freedom, periodic orbits are the 
only generic closed invariant sets in classical phase space besides the 
energy shell. 
Symmetries can introduce other such sets as we have seen above. 
In systems with more than two degrees of freedom a rich structure of 
such subsets may develop. 
Especially in the case of rotationally invariant systems of identical 
particles~\cite{rotinv}, the interplay between the discrete subgroups 
of the rotation group on the one hand and the permutation group on the 
other hand naturally guarantee the existence of low dimensional invariant 
manifolds in classical phase space~\cite{PaSel}. 
Figure~\ref{figure1} shows various configurations that restrict the 
motion to submanifolds for a rotationally invariant system of four 
identical particles. 
Note that configuration~3 can never correspond to central configurations 
in the sense defined at the end of the previous section. 
In higher dimensional subspaces this would be the typical situation.

On these submanifolds the classical motion of the particles is highly
correlated and depends on a small number of "collective" degrees of 
freedom only. 
In practice the summation over the short periodic orbits of submanifolds
depending on one or two collective degrees of freedom is possible.
The search for these orbits as well as the computation of their period
and action may be performed in the appropriate low dimensional
submanifold. 
The computation of their stability exponents and their Maslov index 
however has to take place in full phase space and incorporates the 
many--body effects.

The importance of any invariant submanifold depends crucially on its 
stability properties. 
In what follows we show (i)~that periodic orbits inside such submanifolds 
may be fairly stable and short \cite{PaSel} and thus important for the 
long range correlations in the density of states and (ii)~that entire 
submanifolds may display a classical motion that is weakly unstable in 
transverse directions while being very unstable inside the submanifold.
We show that such a behavior leads to a notable enhancement of the 
revival in the autocorrelation function and thus indicates some degree 
of localization around an invariant submanifold which has collective 
character and is associated with scars \cite{Prosen97,PSW}. 

\subsection{Short periodic orbits inside invariant submanifolds}

We consider in two dimensions a system of $N$ electrons bound by an 
harmonic oscillator.
In polar coordinates the Hamiltonian reads
\be
H = \frac{1}{2}\sum_{i=1}^N \left(p_i^2+\frac{l_i^2}{r_i^2}\right)
+ \sum_{i=1}^N \frac{1}{2}r_i^2
+ \sum_{i<j}\frac{1}{\sqrt{r_i^2+r_j^2-2r_ir_j \cos(\varphi_i-\varphi_j)}}.
\label{ham}
\ee
The simplest periodic orbit may be constructed from the requirement that 
initial conditions should be connected by the group $\mathrm C_{Nv}$, 
i.e. we enter the equations of motion with the ansatz
\be
r_i=r(t), \qquad p_i=p(t), \qquad \varphi_i=2\pi\frac{i}{N},\qquad
\l_i=0, \quad i=1...N.
\ee
Then the time evolution of the functions $r(t)$ and $p(t)$ is governed 
by the Hamiltonian
\be
\tilde{H}= \frac{1}{2}p^2+\frac{1}{2}r^2+\frac{c_N}{r},\qquad 
c_N=\frac{1}{N}\sum_{i<j}
\frac{1}{\sqrt{2-2 \cos \left(2 \pi \frac{i-j}{N}\right)}}.
\ee

\begin{figure}[htb]
\noindent\centerline{
\psfig{figure=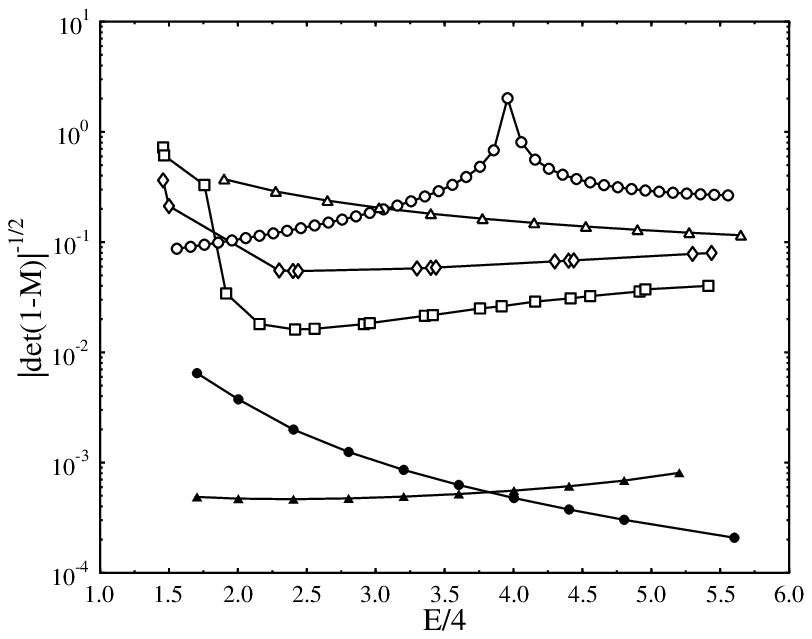,width=0.5\textwidth,angle=0}}
\caption{Stability factor $| det (1-M)|^{-1/2}$ as function of energy 
$E/N$ for $N=4$. (Orbit 1: circles, orbit 2: triangles, orbit 3: diamonds, 
orbit 4: filled diamonds, orbit 5: squares, orbit 6: filled circles, 
orbit 7: filled triangles. 
The orbit--number refers to the corresponding configuration shown 
in Fig.~\protect\ref{figure1})}
\label{monfig}
\end{figure}

At zero total angular momentum and for numbers of particles in the 
range $3 \le N \le 9$ we computed a total of 40 periodic orbits. 
The configurations considered depend at most on two degrees of freedom 
and evolve from small oscillations near equilibrium configurations.
They are shown schematically in Fig.~\ref{figure1} for $N=4$. 
We traced each orbit for a range of energy and computed its period 
$T$ and its full phase space monodromy matrix $M$.
Periods are approximately in the range $1.5<T<4$. 
By comparison with periods of true many--body orbits they are found to 
be among the short ones.
The considered periodic orbits enter into the periodic--orbit sum of
eq.~(\ref{pos}) with a factor proportional to $|det(1-M)|^{-1/2}$, 
where the eigenvalues corresponding to zero Lyapunov exponent due to 
the continuous symmetries of the problem have been eliminated. 
Most periodic orbits are very unstable but for any $N$ we also found a 
few periodic orbits that are stable or only slightly unstable. 
In Fig.~\ref{monfig} a logarithmic plot of $|det(1-M)|^{-1/2}$ versus 
energy is shown for $N=4$.

\subsection{Scars of closed invariant submanifolds}

We consider a system of four particles in two dimensions with Hamiltonian
\be
H = \sum_{i=1}^4\left(\frac{1}{2m}{\bf p}_i^2 
+ 16\alpha|{\bf r}_i|^4\right) 
- \alpha\!\!\sum_{1\le i < j\le 4}\!\!|{\bf r}_i - {\bf r}_j|^4,
\label{fullHam}
\ee
where ${\bf p}_i\!=\!(p_{x_i},p_{y_i})$ and ${\bf r}_i\!=\!(x_i,y_i)$ 
with $i = 1, \ldots, 4$ are the two--dimensional momentum and position
vectors of the $i^{th}$ particle, respectively. 
We use units where $m=\alpha=1, \hbar=0.01$; then coordinates and momenta 
are given in units of $\hbar^{1/3} \alpha ^{-1/6} m^{-1/6}$ and 
$\hbar^{2/3}\alpha^{1/6}m^{1/6}$, respectively. 
The corresponding classical Hamiltonian has the scaling relation
$H(\gamma^{\frac{1}{2}}{\bf p}, \gamma^{\frac{1}{4}}{\bf r})=
\gamma H({\bf p}, {\bf r})$. 
This shows that the structure of classical phase space is independent of 
energy. 
Moreover, energy and total angular momentum are the only integrals of 
motion, and the system is non--integrable. 

We choose the initial conditions in such a way that positions and momenta 
exhibit the symmetry of a rectangle. 
Such a configuration is shown in Fig.~\ref{figure1}, configuration 3. 
The appropriate manifold is spanned by the two--dimensional vectors 
${\bf p}=(p_x,p_y)$ and ${\bf r}=(x,y)$ giving the momentum and position 
of particle~1. 
The associated Hamiltonian 
$\tilde{H}({\bf p},{\bf r})=\frac{1}{2}{\bf p}^2 + 16x^2y^2$ has been 
studied extensively in the literature, both classical and the quantum 
mechanics~\cite{qhamil}.
The classical system is essentially chaotic. 
One very small island of stability is known~\cite{Dahl90}. 
The stability exponents of periodic orbits in the central region of the 
manifold are quite large. 
However, the picture changes when one considers the stability in the 
directions transverse to the submanifold. 
The sums of the transverse stability exponents of the central orbits are 
found to be considerably smaller than the stability exponents inside the 
manifold. 
This shows that a trajectory may be trapped quite a time in the vicinity 
of the submanifold before leaving it.
One might therefore expect that this entire manifold rather than a specific 
periodic orbit may scar wave functions of the corresponding quantum system.

To demonstrate this conjecture we consider the time evolution of a 
Gaussian wave packet
\be
\Psi({\bf r},t)=c\exp\left[-\frac{1}{2}({\bf r}-{\bf r}_0)^TA({\bf r}-
{\bf r}_0)+\frac{i}{\hbar}{\bf p}_0^T({\bf r}-{\bf r}_0)\right]
\ee
where $<\bf p>= {\bf p}_0$ and $<\bf r>={\bf r}_0$ define a point on the 
submanifold. We have used the shorthand notation
${\bf r}=({\bf r}_1,{\bf r}_2,{\bf r}_3,{\bf r}_4)$ and
${\bf p}=({\bf p}_1,{\bf p}_2,{\bf p}_3,{\bf p}_4)$
for configuration and momentum space vectors, respectively.
The autocorrelation function $C(t)=<\Psi(t=0)| \Psi(t)>$ is computed
in the semi-classical approximation. 
Within the manifold, we used Heller's cellular 
dynamics~\cite{Hellcell,SapHel} which takes into account the nonlinearity 
of the classical motion. In the transverse direction we used linearized 
dynamics. This approximation is justified in the time scales considered, 
since the classical return probability to the manifold of transversely
escaping trajectories is negligible. 

\begin{figure}[htb]
\noindent\centerline{
\psfig{figure=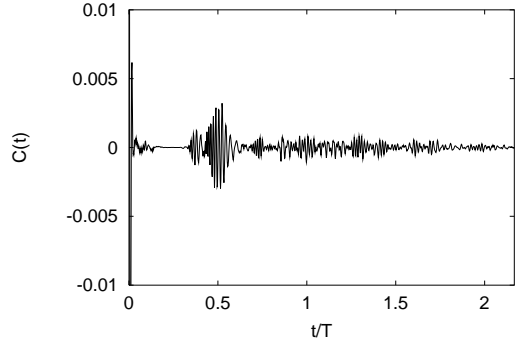,width=0.5\textwidth,angle=270}}
\caption{Autocorrelation function $C(t)$ of a symmetrized wave packet
launched on a periodic orbit with period $T$ inside a weakly unstable
manifold. In addition to the linear revival around $t=\frac{T}{2}$, 
strong nonlinear revival is seen for larger times.} 
\label{Fig2a}
\end{figure}

We launch wave packets along periodic or aperiodic orbits within the 
submanifold and consider their revival as measured by the autocorrelation
function. 
To achieve shorter recurrence times the initial packet was symmetrized 
with respect to the reflection symmetry of the system within the 
submanifold.
For not too unstable periodic orbits we expect a fairly strong revival
after one period, known as the linear revival~\cite{SapHel}. 
As an example, we show in Fig.~\ref{Fig2a} the real part of the 
autocorrelation function starting on the shortest periodic orbit with 
period $T$ inside the manifold. 
We indeed find strong linear revival.
However, at larger times we find scattered strong revivals, the revival
corresponding to the second period not being dominant. 
This implies that a significant fraction of the original amplitude remains 
within the submanifold, and that this fact is not related to the periodic 
orbit we started on. 
Revivals calculated for packets started on aperiodic orbits show similar 
features except for the obvious absence of the linear revival.

Numerical studies show that the configurations~(6) and~(7) of 
Fig.~\ref{figure1} belong to two dimensional submanifolds that are very 
unstable in the transverse directions. 
Launching wave packets inside this manifold does not lead to any 
significant linear or nonlinear revival. 

In summary, we have seen that invariant submanifolds of rotationally 
invariant systems of identical particles constitute a remarkable structure 
in classical phase space. 
On such manifolds short periodic orbits may be easily found. 
Furthermore a chaotic submanifold with small transverse stability exponents 
may trap classical trajectories quite a time in their vicinity and lead 
to scaring of wave functions. 
As such submanifolds involve motion of many particles in a low-dimensional 
space we may hope that they constitute an approach to collective motion 
in chaotic systems.

\section{Beyond symmetric Hamiltonians}

At last let us turn to an interesting observation that extends the 
usefulness of symmetric periodic orbits even to non-symmetric systems. 
Consider the usual symmetric three-disc scattering system with sufficient 
distance between the discs to obtain the usual binary 
dynamics~\cite{threedisc}. 
In this case we have a ${\mathrm C}_{3v}$ symmetry and five fundamental 
orbits, three bouncing between two discs and two cyclic ones with opposite 
directions. 
Cyclic symmetry makes the three bouncing orbits equivalent and reflection 
symmetry the two cyclic ones.

It is easy to see that this fits with our expectations of the 
importance of symmetric orbits. 
A periodic orbit with the full ${\mathrm C}_{3v}$ symmetry does not exist. 
So we have to consider the subgroups. 
What is more surprising~\cite{JungSel} is that the structural stability of 
a complete horseshoe guarantees that these same orbits will be the 
fundamental ones even if we distort the system a little {\it e.g.} by 
choosing different sizes for the discs and/or distort the equilateral 
triangle on which the centers of the discs are placed.
Inspection of the specific system shows that if the discs are sufficiently 
far apart as compared to their radii the deformations can become fairly 
large because the binary dynamics will only break down if one of the discs 
will interfere with an orbit connecting the other two.

Conversely we can ask the question whether we can construct fundamental 
orbits that do not have the corresponding symmetries.
For this purpose we shall look at a the development of a two hill system 
into a four hill system.

We consider the Hamiltonian 
\ba
H&=&{1\over 2}(p_x^2 + p_y^2)+\exp[-(x-a)^2-(y-A)^2] 
 + \exp[-(x+a)^2-(y-A)^2] \nonumber \\
&+& \exp[-(x-b)^2-(y+A)^2] + \exp[-(x+b)^2-(y+A)^2],
\label{hills}
\ea
where $ A$ is the fixed separation of the original two hills. 
For $a = b $ this system has a symmetry group of order 4 generated by two 
reflections, one along the axis uniting the two hills (y-axis) and one 
perpendicular to this axis at the center between the two hills (x-axis). 
Now we restrict our considerations to the single energy value $E=0.5$ and 
start the development scenario at $a=b=0.7$ where the equipotential line 
$V=0.5$ still consists of two convex components. 
Then the whole invariant set consists of the single unstable periodic 
trajectory which oscillates along the y-axis between the two potential 
mountains. 
Despite of the fact that this system is integrable the invariant set
is hyperbolic and by consequence structurally stable against small 
deformations.

\begin{figure}
\noindent\centerline{
\psfig{figure=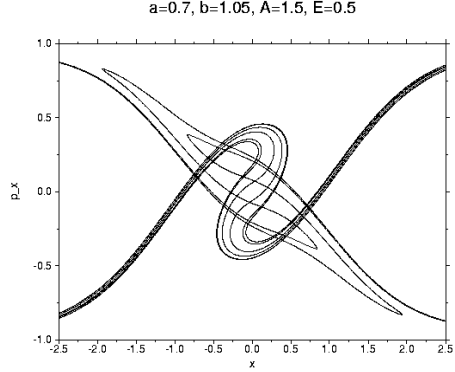,height=5cm,angle=270}}
\caption{
Complete ternary horseshoe formed by the invariant manifolds of the two 
outer unstable fixed points. The Poincar\'e map is taken in the surface 
$y=0$ and $E=0.5$. The parameter values are $a=0.7$ and $b=1.05$.}
\label{ternary}
\end{figure}

Keeping $ a=0.7 $ fixed and varying $ b $ the symmetry of the system is 
reduced to a simple reflection symmetry in the y-axis. 
The original hyperbolic periodic orbit undergoes a pitchfork bifurcation 
(at $b \approx 0.73$) to one elliptic orbit along the symmetry line 
(y-axis) and a pair of non symmetric hyperbolic orbits that are reflection
images of each other. 
As we increase $ b $ further the elliptic orbit undergoes a period 
doubling bifurcation and becomes inverse hyperbolic. 
The horseshoe develops to a complete ternary one for $ b =1.05 $ shown in 
Fig.~\ref{ternary}.

\begin{figure}
\noindent\centerline{
\psfig{figure=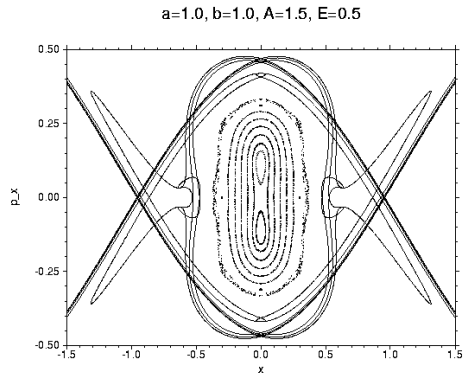,height=5cm,angle=270}}
\caption{
Incomplete horseshoe with five fixed points in the same surface of section 
as in previous figure. 
We show the invariant manifolds of the two outer fixed points, several 
KAM islands in the region of stability around the three inner fixed 
points and one small chaotic strip.}
\label{orderfive}
\end{figure}

If we keep $b$ fixed at $b=1.0$ and increase $ a $ starting from $0.7$
the horseshoe of order three reduces its degree of development but at 
some point the central orbit undergoes a further pitchfork bifurcation 
for fixed $ b = 1.0$ when  $ a \approx 0.97$. 
Note that this value of $ a $ is quite near to the value of $ b $. 
If we further increase $ a $ such that $ a = b = 1.0 $ we have a 
symmetric horseshoe of order five as shown in Fig.~\ref{orderfive}.
In this case we are back to the typical situation where all fundamental 
periodic orbits are symmetric under a subgroup of the symmetry group 
which is again generated by two reflections taking pairs of hills into 
each other. 
This shows that the five orbits we found after the bifurcation are 
again simple deformations of symmetric ones.

If instead of increasing $ a $ we further increase $ b $ the central 
orbit will undergo a further pitchfork bifurcation generating two orbits 
with opposite rotational orientation in order finally to become a homoclinic
connection of a new periodic orbit that is formed on the saddle 
perpendicular to the former central orbit. 
At this point we have a three hill problem which upon sufficient separation 
of the hills will have a horseshoe of the type discussed at the beginning 
of this section for the three disc system.  

\begin{figure}
\noindent\centerline{
\psfig{figure=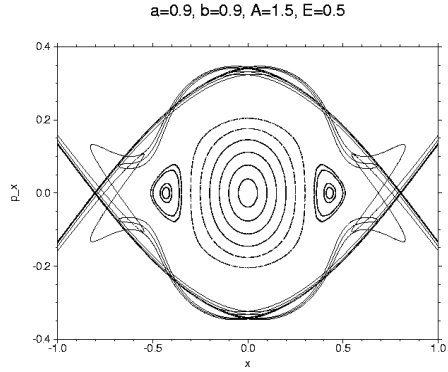,height=5cm,angle=270}}
\caption{
Incomplete ternary horseshoe in the same surface of section as in previous 
figures. We show the invariant manifolds of the two outer fixed points, 
several KAM islands in the region of stability around the inner fixed 
point and a secondary structure of order two.}
\label{nopitchfork}
\end{figure}

It thus seems that we have found that the horseshoe of order three really 
has orbits that are not deformations of symmetric ones. 
This appearance is fallacious because we can take a different route of 
deformation. 
Choosing the value of $b=0.9$, { \it i.e.} sufficiently small we can vary 
$a $ up to $ a = b $ without causing the pitchfork bifurcation of the 
central orbit as we can see in Fig.~\ref{nopitchfork}.
The fact that the picture is not hiding a very small splitting of the 
central orbit is readily checked by verifying that it is still inverse 
hyperbolic.

The question whether the fundamental periodic orbits of a chaotic problem 
can always be viewed as deformations of those orbits symmetric under a 
non-trivial subgroup of the symmetry group of a more symmetric problem 
remains an open question.
That it will be often the case is quite evident. 
As we know from refs.~\cite{Henon1,richter}, there are particularly 
efficient methods to find symmetric periodic orbits. Therefore it 
might in certain cases be well worth considering such a deformation 
to determine the structure of a system without or with low symmetry.

\section{Summary and conclusions}

A modern analysis of the chaotic saddle of the Copenhagen problem proves
a thirty year old statement of H\'enon to be right from a novel point of 
view. 
The simple (according to H\'enon) periodic orbits are sufficient 
to explain the horseshoe structure of the entire chaotic saddle in terms 
of their stable and unstable manifolds. 
The only addition that is necessary are the symmetric parabolic orbits 
which are the limits of various families of symmetric periodic orbits and 
which generate the edges of what is known as the fundamental rectangle. 

The concept of symmetry generated invariant submanifolds presents a
particular challenge in this context: The periodic (and incidentally 
also the aperiodic) orbits that lie within these manifolds are all
symmetric in a very special way. 
The symmetry transforms such orbits pointwise into each other.
Such manifolds provide very easily some of the periodic orbits we may need 
to describe our system. 
We furthermore note that in the context of semi-classics these manifolds 
may be of fundamental importance.
Actually there are strong signs that such manifolds may carry collective 
states with fairly long life times.

Finally we use the structural stability to argue that symmetric horseshoes 
may be readily deformed into non-symmetric ones and vice versa. 
This implies that the topological structure of the fundamental periodic 
orbits does not change over some finite interval for the parameter that 
measures the symmetry breaking deformation of our system. 
Thus the symmetry even governs systems in which it is broken.

The remarkable and seemingly entirely uncorrelated roles of symmetric 
orbits in the systems discussed in sections~2 and~3 as well as others
given in the literature~\cite{richter,Gutz,Wint,cvit} is the central 
question of our considerations.
The survival of basic symmetric orbits in the horseshoe construction for
deformed systems gives an indication of the origin of their importance.

Finally we wish to emphasize that the entire paper raises more questions 
than it answers, and indeed it is meant to be a stimulus for researchers 
in the field not to take the common knowledge for granted, but rather 
evaluate the situation in their specific fields of interest.


\begin{thebibliography}{99}

\bibitem{Henon1} 
M. H\'enon, {\it Ann. Astrophys.} {\bf 28} (1965) 499.

\bibitem{Henon3} 
M. H\'enon, {\it Bull. Astron.} (3) {\bf 1} fasc 1 (1966) 57; 
fasc 2 (1966) 49.

\bibitem{richter} 
P.H. Richter, H.J. Scholz and A. Wittek, {\it Nonlinearity} {\bf 3} 
(1990) 45.

\bibitem{Gutz} 
M.C. Gutzwiller, {\em Chaos in Classical and Quantum Mechanics},
 (Springer, Berlin) 1990; {\it J. Math. Phys.} {\bf 12} (1971) 343.

\bibitem{Wint}
 D. Wintgen, K. Richter and G. Tanner, {\it Chaos} {\bf 2} (1992) 32. 

\bibitem{cvit} 
P. Cvitanovi\'c and B. Eckhardt, {\it Nonlinearity} {\bf 6} (1993) 277.

\bibitem{JungRuck1} 
B. R\"uckerl, C. Jung, {\it J. Phys. A} {bf 27} (1994) 55.

\bibitem{celmec1} 
L. Benet, et al., {\it Celest. Mech. \& Dyn. Astr.} {\bf 66} (1997) 203.

\bibitem{celmec2}
L. Benet, et al., {\it Celest. Mech. \& Dyn. Astr.}, submitted.

\bibitem{centralconf}
A. Wintner, {\em The Analytical Foundations of Celestial Mechanics}
 (Princeton University Press, 1941);
D.G. Saari, {\it Celest. Mech. \& Dyn. Astr.} {\bf 21} (1980)~9.

\bibitem{Prosen97} 
T. Prosen, {\it Phys. Lett. A} {\bf 233} (1997)~332.

\bibitem{PaSel} 
T. Papenbrock and T.H. Seligman, {\it Phys. Lett. A} {\bf 218} 
(1996)~229.

\bibitem{PSW}
 T. Papenbrock, T.H. Seligman, and H.A. Weidenm\"uller 
{\it Phys. Rev. Lett.} {\bf 80} (1998)~3057

\bibitem{potheory}
 R. Balian and C. Bloch, {\it Ann. Phys.} (N.Y.) {\bf 69} (1972)~76;
 M.V.Berry, in {\em Chaotic Behavior of Deterministic Systems},
 edited by G. Iooss {\em et al.} (North--Holland, Amsterdam, 1983).

\bibitem{fewdegrees}
 D. Delande and J.C. Gay,  {\it Phys. Rev. Lett.} {\bf 57} (1986)~2006;
 M.C. Du and J.B. Delos, {\it Phys. Rev. Lett.} {\bf 58} (1987) 1731;
 See {\em Chaos and Quantum Physics}, eds. M.-J. Giannoni {\em et al.}
  (North--Holland, Amsterdam, 1991) and references therein.

\bibitem{scars} 
 E.J. Heller, {\it Phys. Rev. Lett.} {\bf 53} (1984) 1515;
 E.B. Bogomolny, {\it Physica} {\bf D 31} (1988) 169;
 M.V. Berry, {\it Proc. Roy. Soc. Lond.} {\bf A 423} (1989) 219;
 O. Agam and S. Fishman, {\it J. Phys. A} {\bf 26} (1993) 2113, 
{\it Corrigendum} 6595; {\it Phys. Rev. Lett.} {\bf 73} (1994) 806.

\bibitem{rotinv}
 M.H. Sommermann and H.A. Weidenm\"uller, {\it Europhys. Lett.} {\bf 23}
  (1993) 79;
 H.A. Weidenm\"uller, {\it Phys. Rev.} {\bf A48} (1993) 1819;
 T.H. Seligman and H.A. Weidenm\"uller, {\it J. Phys. A} {\bf 27} 
(1994) 7915.

\bibitem{qhamil} 
 B. Simon, {\it Ann. Phys.} {\bf 146} (1983) 209;
 C.C. Martens, R.L. Waterland and W.P. Reinhardt, {\it J. Chem. Phys.} 
{\bf 90} (1989) 2328;
 S. Tomsovic,  {\it J. Phys. A} {\bf 24} (1991) L733;
 P. Dahlqvist and G. Russberg, {\it J. Phys. A} {\bf 24} (1991) 4763;
 O. Bohigas, S. Tomsovic, D. Ullmo, {\it Phys. Rep.} {\bf } (1993) 43;
 B. Eckhardt, G. Hose, E. Pollak, {\it Phys. Rev.} {\bf A 39} (1989) 3776.

\bibitem{Dahl90}
 P. Dahlqvist and G. Russberg, {\it Phys. Rev. Lett.} {\bf 65} (1990) 2837.

\bibitem{Hellcell}
 E.J. Heller, {\it J. Chem. Phys.} {\bf 94} (1991) 2723.

\bibitem{SapHel}
 M.A. Sepulveda, E.J. Heller, {\it J. Chem. Phys.} {\bf 101} (1994) 8004.

\bibitem{threedisc} 
 B. Eckhardt, {\it J. Phys. A} {\bf 20} (1987) 5971;
 P. Gaspard and S. Rice, {\it J. Chem. Phys.} {\bf 90} (1989) 2225.

\bibitem{JungSel}
 C. Jung, C. Lipp and T.H. Seligman, {\it Ann. Phys.} (N.Y.) submitted.

\end{thebibliography}
\end{document}